\documentclass[%
 reprint,
superscriptaddress,
showpacs,preprintnumbers,
 amsmath,amssymb, 
 aps,
 notitlepage,
]{revtex4-1}

\usepackage{graphicx}% Include figure files
\usepackage{dcolumn}% Align table columns on decimal point
\usepackage{bm}% bold math
\usepackage{float}% float
\usepackage{hyperref}% add hypertext capabilities
\usepackage[left]{lineno}

\newcommand{\ks}{K_{S}^{0}}
\newcommand{\kkjpsi}{K\bar{K}J/\psi}
\newcommand{\kpkmjpsi}{K^{+}K^{-}J/\psi}
\newcommand{\ksksjpsi}{\ks\ks J/\psi}
\newcommand{\pippimjpsi}{\pi^{+}\pi^{-}J/\psi}
\newcommand{\pipijpsi}{\pi\pi J/\psi}
\newcommand{\ecm}{E_{\mathrm{CM}}}
\newcommand{\sige}{\sigma}
\newcommand{\mll}{\mathrm{M}(l^+l^-)}
\newcommand{\mkpkm}{\mathrm{M}(K^+K^-)}

\begin{document}

\title{Observation of $e^{+}e^{-} \rightarrow K\bar{K}J/\psi$ at center-of-mass energies from 4.189 to 4.600~GeV}

\author{
M.~Ablikim$^{1}$, M.~N.~Achasov$^{9,d}$, S. ~Ahmed$^{14}$, M.~Albrecht$^{4}$, A.~Amoroso$^{50A,50C}$, F.~F.~An$^{1}$, Q.~An$^{47,39}$, J.~Z.~Bai$^{1}$, O.~Bakina$^{24}$, R.~Baldini Ferroli$^{20A}$
% ...
}

\affiliation{
~Institute of High Energy Physics, Beijing 100049, People's Republic of China\\
% ...
$^{4}$ Bochum Ruhr-University, D-44780 Bochum, Germany\\
% ...
$^{9}$ G.I. Budker Institute of Nuclear Physics SB RAS (BINP), Novosibirsk 630090, Russia\\
% ...
$^{14}$ Helmholtz Institute Mainz, Johann-Joachim-Becher-Weg 45, D-55099 Mainz, Germany\\
% ...
$^{20}$ (A)INFN Laboratori Nazionali di Frascati, I-00044, Frascati, Italy; (B)INFN and University of Perugia, I-06100, Perugia, Italy\\
% ...
$^{24}$ Joint Institute for Nuclear Research, 141980 Dubna, Moscow region, Russia\\
% ...
$^{39}$ State Key Laboratory of Particle Detection and Electronics, Beijing 100049, Hefei 230026, People's Republic of China\\
% ...
$^{47}$ University of Science and Technology of China, Hefei 230026, People's Republic of China\\
% ...
$^{50}$ (A)University of Turin, I-10125, Turin, Italy; (B)University of Eastern Piedmont, I-15121, Alessandria, Italy; (C)INFN, I-10125, Turin, Italy\\
% ...
$^{d}$ Also at the Novosibirsk State University, Novosibirsk, 630090, Russia\\
% ...
}
 \collaboration{BESIII Collaboration}
 \noaffiliation

\begin{abstract}
We investigate the process $e^{+}e^{-} \rightarrow \kkjpsi$
at center-of-mass energies from 4.189 to 4.600~GeV
using 4.7~fb$^{-1}$
of data collected by the BESIII detector at the BEPCII collider.
The Born cross sections for the reactions $e^{+}e^{-} \rightarrow \kpkmjpsi$ and $\ksksjpsi$ are measured as a function of center-of-mass energy.  The energy dependence of the cross section for $e^{+}e^{-} \rightarrow \kpkmjpsi$ is shown to differ from that for $\pippimjpsi$ in the region around the $Y(4260)$.
In addition, there is evidence for a structure around 4.5~GeV in the $e^{+}e^{-} \rightarrow \kpkmjpsi$ cross section that is not present in $\pippimjpsi$.

\end{abstract}


%\begin{linenumbers}

\maketitle

The $Y(4260)$ resonance was first discovered in the process $e^{+}e^{-} \rightarrow Y(4260) \rightarrow \pi^{+}\pi^{-}J/\psi$ by the BaBar experiment~\cite{BABAR} using the initial state radiation~(ISR) technique and then later confirmed by CLEO~\cite{CLEO} and Belle~\cite{Belle}.  This state does not fit into the conventional charmonium spectrum of the quark model~\cite{Barnes}, which predicts three vector charmonium states in this mass region, usually identified as the experimentally established $\psi(4040)$, $\psi(4160)$, and $\psi(4420)$ 
states~\cite{PDG}.
In addition, even though the mass of the $Y(4260)$ is well above the open-charm $D\bar{D}$ threshold, it has not yet been found to decay to $D\bar{D}$~\cite{opencharm}, in contrast to the conventional charmonium states in this mass region.
There are several theoretical interpretations of the $Y(4260)$, including tetraquark~\cite{tetraquark}, meson molecule~\cite{DDmole}, hadroquarkonium~\cite{hadroquark}, hybrid meson~\cite{hybrid}, and others~\cite{y4260others}.

In addition to $e^{+}e^{-} \rightarrow \pippimjpsi$, the $Y(4260)$ state has been searched for in many other modes, including $\pi\pi h_{c}$~\cite{pipihc,pizpizhc}, $\omega\chi_{cJ}$~\cite{omegachic}, $\eta J/\psi$~\cite{etajpsi,belleetajpsi},  
$\eta' J/\psi$~\cite{etaprime} and $\kkjpsi$~\cite{kkjpsiBelle}.  
Rather than showing conclusive evidence for new $Y(4260)$ decay modes, the energy dependencies of the $e^+e^-$ cross sections hint at a more complex pattern than just the existence of a $Y(4260)$. 
More recent results from BESIII, in the $\pi^{+}\pi^{-}J/\psi$\cite{pipijpsi2} and $\pi^{+}\pi^{-}h_{c}$\cite{pipihc2} final states, show two resonant structures within this region.  
In order to understand this mass region, it is thus important to measure additional $e^+e^-$ cross sections.  In particular, measuring the ratio of $\kkjpsi$ and $\pipijpsi$ cross sections would allow us to gain new insight into the nature of the $Y(4260)$~\cite{Theory-CF}.
  
In the following, we use 4.7~fb$^{-1}$ of data collected at the Beijing Spectrometer~(BESIII) with center-of-mass energies~($\ecm$) ranging from 4.189 to 4.600~GeV to measure the Born cross sections~$(\sige)$ of the reactions $e^{+}e^{-} \rightarrow \kpkmjpsi$ and $\ksksjpsi$.  To identify whether or not the $\kpkmjpsi$ system originates from a $Y(4260)$, the energy dependence of the $e^{+}e^{-}\to \kpkmjpsi$ cross section is compared to that of $\pippimjpsi$.  
The ratio $\sige(\ksksjpsi)$/$\sige(\kpkmjpsi)$ is also calculated to test isospin symmetry.

The BESIII experiment uses a general purpose magnetic spectrometer \cite{BESIIIHardware}.  A superconducting solenoid magnet provides a 1.0~T field, enclosing a helium-gas-based drift chamber~(MDC) for charged particle tracking, a plastic scintillator time-of-flight system~(TOF) for particle identification~(PID), and a CsI(Tl) Electromagnetic Calorimeter~(EMC) to measure the energy of neutral particles.  The Beijing Electron Positron Collider~(BEPCII) uses two rings to collide electrons and positrons with $\ecm$ from 2.0 to 4.6~GeV.

The data samples used in this analysis were collected at 14 different $\ecm$~\cite{BES_ECM}.  
Large data sets were collected at 
4.226~(1092~pb$^{-1}$), 
4.258~(826~pb$^{-1}$), 
4.358~(540~pb$^{-1}$), 
4.416~(1074~pb$^{-1}$), 
4.467~(110~pb$^{-1}$), 
4.527~(110~pb$^{-1}$), and
4.600~(567~pb$^{-1}$)~GeV.  
Other smaller samples of 50~pb$^{-1}$ each
were collected at
4.189, 
4.208, 
4.217, 
4.242,  
4.308,  
4.387,  and
4.575~GeV~\cite{BES_Lumin}.

\textsc{geant4}-based~\cite{Geant4} Monte Carlo~(MC) simulations are used to study efficiencies and backgrounds.  Signal MC samples are generated for $e^{+}e^{-} \to \pippimjpsi$, $\kpkmjpsi$, and $\ksksjpsi$ using \textsc{evtgen}~\cite{EvtGen} and assuming a phase space model for all decays.  \textsc{kkmc}~\cite{KKMC} is used to calculate the ISR correction factors needed to convert an observed cross section to a Born cross section~\cite{ISR,prg}.

\begin{figure}[t]
\centerline{\includegraphics[width=0.80\columnwidth]{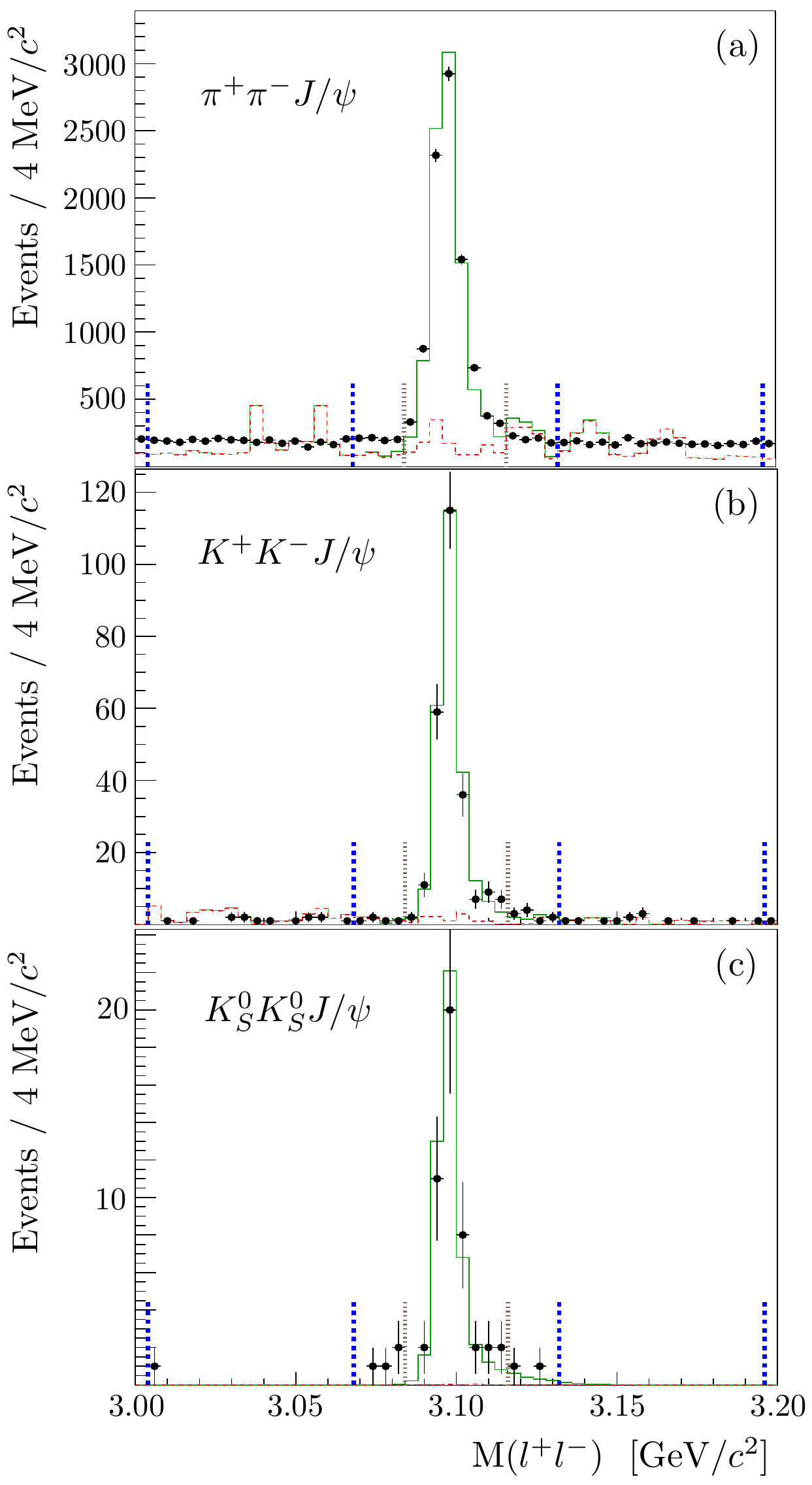}}
\caption{\label{fig:jpsimass}
(Color online) The distribution of lepton pair mass, $\mll$, for
(a)~$\pippimjpsi$, (b)~$\kpkmjpsi$, and (c)~$\ksksjpsi$. Data from all $\ecm$ are combined.  Points are for data; the green solid histograms are for signal MC events; and the red dashed histograms are for background MC events. The signal regions are shown by the gray dashed lines, while the sideband regions are shown with the blue dotted lines.}
\end{figure}

Background MC samples are divided into three categories: quantum electrodynamic~(QED), continuum, and peaking backgrounds.  
%For QED, MC samples are simulated at 4.23, 4.26, 4.36, 4.42, and 4.60~GeV separately.  
For the continuum backgrounds, samples are generated for $e^{+}e^{-} \to 4\pi$, $6\pi$, $2K2\pi$, $2K4\pi$, and $p\bar{p}\pi\pi$.  The cross sections for these channels were measured separately and were found to be on the order of 100~pb. For the peaking backgrounds, where a background $J/\psi$ may be present, samples are generated for 
$e^{+}e^{-} \to \eta J/\psi, \eta' J/\psi, \pi^{+}\pi^{-}\psi(3686)$, 
and $\pi^{0}\pi^{0}\psi(3686)$ according to their known cross sections~\cite{etajpsi,etaprime,pipipsi2S}.  Other sources of backgrounds, including those from ISR or $D\bar{D}$, are also generated and are found to be negligible.

Final states in this analysis include $\kpkmjpsi$ and $\ksksjpsi$, where the $J/\psi$ decays into $e^{+}e^{-}$ or $\mu^{+}\mu^{-}$, and each $\ks$ decays into $\pi^{+}\pi^{-}$.  In addition, the previously studied final state of $\pippimjpsi$~\cite{BESIII, pipijpsi2} is reconstructed to cancel systematic uncertainties when calculating ratios of cross sections.

To select events, we require at least two positively charged and two negatively charged tracks for the $\kpkmjpsi$ and $\pippimjpsi$ modes and at least three positively charged and three negatively charged tracks for the $\ksksjpsi$ mode.  If more than one combination passes the selection, multiple counting of events is allowed.  However, our selection removes all significant combinatoric backgrounds, according to studies of the MC samples.  A distance of closest approach for any primary charged track from the beam interaction point must be within $\pm$10~cm along the beam direction, and 1~cm in the plane perpendicular to the beam direction.  The polar angle in the MDC for each charged track must satisfy $|\cos(\theta)|< 0.93$.  To identify leptons, the energy deposited in the calorimeter divided by the momentum of any lepton candidate must be greater than 0.80 for either electron or less than 0.25 for both muons.

We perform a four-constraint~(4C) kinematic fit for $\pippimjpsi$ and $\kpkmjpsi$ and a six-constraint~(6C) fit for $\ksksjpsi$.  For the 4C fits, the four-momentum is constrained to the initial center-of-mass system.  For the 6C fits, the masses of the two $\ks$ are also constrained.  The resulting $\chi^{2}$/dof is required to be less than 10.

To remove radiative Bhabha background events, where the radiated photon converts into an $e^{+}e^{-}$ pair when interacting with the material inside the detector, all pairs of oppositely charged tracks must have an opening angle satisfying $\cos(\theta)<0.98$.  For PID, the TOF and ionization energy loss~($dE/dx$) from the MDC are combined to calculate probabilities for kaon and pion hypotheses of each track.  The charged kaons in $\kpkmjpsi$ are selected by requiring Prob($K$) $>$ Prob($\pi$).  %This selection removes $85.5\%$ of the continuum backgrounds while keeping $98.2\%$ of the predicted signal.  
In the $\ksksjpsi$ channel, in order to remove backgrounds from $e^{+}e^{-} \rightarrow \pi\pi \psi(3686)$ with $\psi(3686)$ decaying to $\pippimjpsi$, each $\ks$ must have $L/\sigma>4$, where $L$ is the $\ks$ decay length and $\sigma$ is its uncertainty.  
The $\pi^{+}$ and $\pi^{-}$ pair from the $\ks$ decay is required to have an invariant mass between 471 and 524~MeV/$c^{2}$ and
originate from a common vertex by requiring the $\chi^{2}$ of a vertex fit be less than 100.

After the above selection, the distributions of dilepton invariant mass, $\mll$, for the three different decay modes (with all 14~$\ecm$ combined) are shown in Fig.~\ref{fig:jpsimass}.  Clear $J/\psi$ signals are observed. Backgrounds outside of the $J/\psi$ signal region are well described by our background MC simulation and are flatly distributed.  For $\pippimjpsi$, the main background is from the process $e^{+}e^{-} \rightarrow \pi^{+}\pi^{-}\pi^{+}\pi^{-}$.  For $\kpkmjpsi$, the main background is from $e^{+}e^{-} \rightarrow K^{+}K^{-}\pi^{+}\pi^{-}$.  
There are no significant peaking background events expected in any mode, with the largest estimated to be $0.4$ events in the $\ksksjpsi$ channel from $e^{+}e^{-} \rightarrow\pi^{+}\pi^{-}\psi(3686)$ $(\rightarrow \pi^{+}\pi^{-}\pi^{+}\pi^{-}J/\psi)$.

\begin{figure}[ht]
\centerline{\includegraphics[width=0.80\columnwidth]{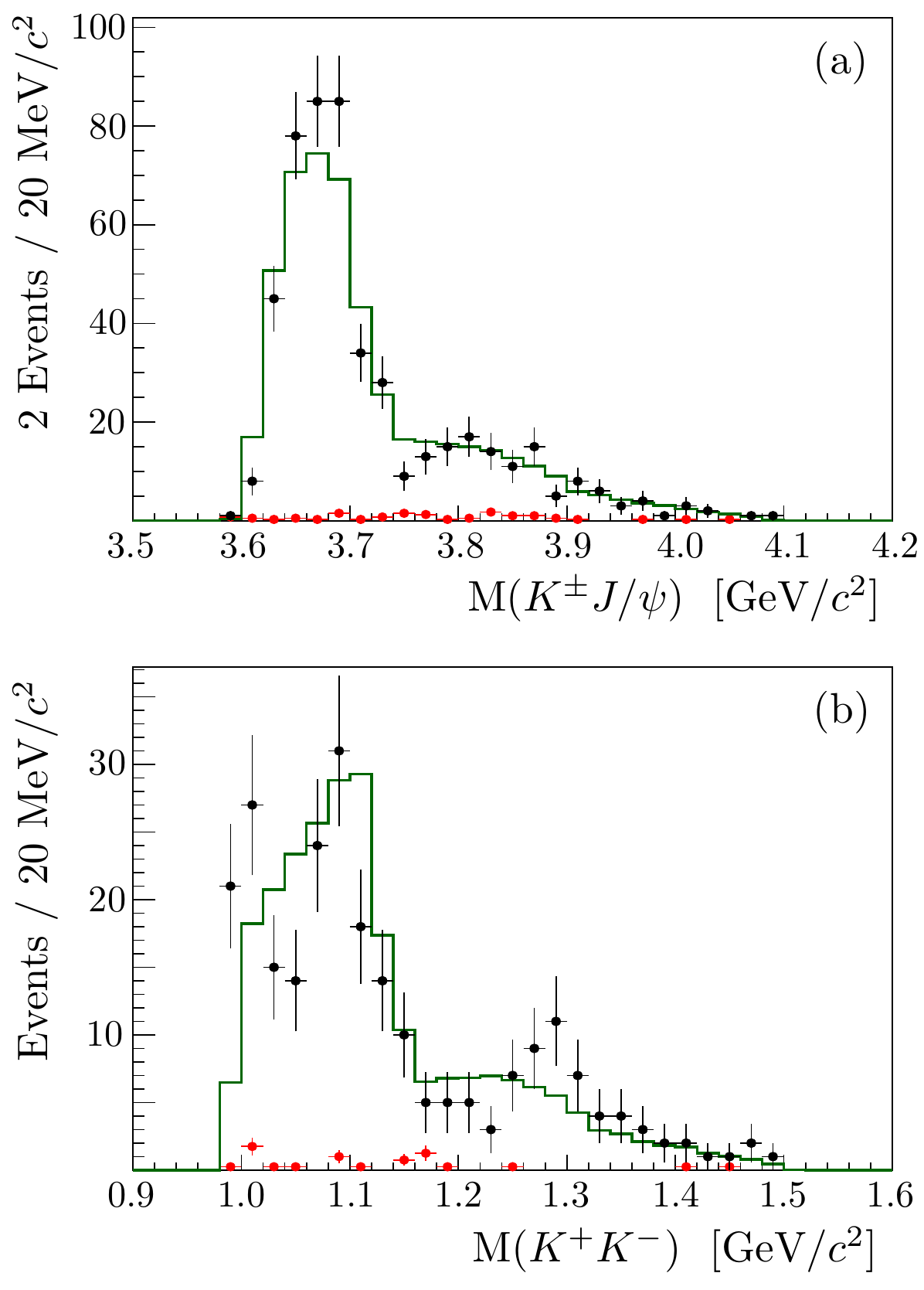}}
\caption{\label{fig:intermed_mass}
(Color online) The invariant mass distributions for (a)~$K^{\pm}J/\psi$~(two entries per event) and (b)~$K^{+}K^{-}$.  Data from all $\ecm$ are combined. Black points are for data from the $J/\psi$ signal region;  red points are for data from the $J/\psi$ sideband regions (normalized to the size of the signal region); dark green solid histograms are for signal MC events~(normalized using the measured cross section at each $\ecm$).}
\end{figure}

To explore potential intermediate states in the $\kkjpsi$ channel, data are compared with phase-space signal MC events in Fig.~\ref{fig:intermed_mass}.  The signal MC histograms are normalized to the measured Born cross section at each $\ecm$. There is no significant difference between the data and MC simulation in the $K^{\pm}J/\psi$ mass distributions.  There are, however, apparent differences in the $K^{+}K^{-}$ invariant mass, where there may be hints of $f_{0}(980)$ and $f_{2}(1270)$ signals.  However, there are not sufficient data to investigate further.

The Born cross section at each $\ecm$ is calculated by:
\begin{equation}
\sige = \frac{N^{\mathrm{sig}}}{\mathcal{L}\epsilon(1+\delta)(1+\delta_{VP})B(J/\psi\rightarrow l^{+}l^{-})}.
\end{equation}
The signal yield, $N^{\mathrm{sig}}$, is calculated by subtracting the number of $J/\psi$ sideband events from the number of $J/\psi$ signal events.  The $J/\psi$ signal region is $3084<\mll<3116$~MeV/$c^2$; and the low and high sideband regions are   $3004<\mll<3068$~MeV/$c^2$ and 
$3132<\mll<3196$~MeV/$c^2$, respectively.  
Uncertainties on the number of signal events are calculated using the Rolke method~\cite{Rolke}.  The total signal yields for all $\ecm$ are $7984^{+99}_{-98}$ events for the $\pippimjpsi$ channel, 
$238^{+16}_{-15}$ for $\kpkmjpsi$, and 
$46.5^{+7.3}_{-6.6}$ for $\ksksjpsi$.
The integrated luminosity values, $\mathcal{L}$, are taken from Ref.~\cite{BES_Lumin}.  The branching fraction
$B(J/\psi \to l^{+}l^{-}) = (11.93 \pm 0.06)\%$
is taken from the particle data group~(PDG)~\cite{PDG}.  
For the $\ksksjpsi$ mode, a factor of 
$B(\ks\rightarrow \pi^{+}\pi^{-})^{2}$ = $(47.9 \pm 0.03)\%$ 
is also included.  
The vacuum polarization factors, $(1+\delta_{VP})$, are taken from Ref.~\cite{Vac_pol}.  The efficiencies for each mode, $\epsilon$, are derived from the signal MC samples incorporating ISR effects.  For $\pippimjpsi$, the efficiencies (without ISR effects) at each energy point are around 48\%.  For $\kpkmjpsi$, the efficiencies range from 13\% at low $\ecm$ to 35\% at high $\ecm$.  For $\ksksjpsi$, the efficiencies are about 25\%.

The ISR correction factors, $(1+\delta)$, are calculated using an iterative procedure.  A cross section following a Breit-Wigner line shape with  PDG values for the mass and width of the $Y(4260)$ is used as the first input for both the $\pippimjpsi$ and $\kkjpsi$ channels.  The resulting cross section line shapes are used as the next inputs, and this procedure is iterated until the Born cross section converges.

\begin{figure}[ht]
\centerline{\includegraphics[width=0.77\columnwidth]{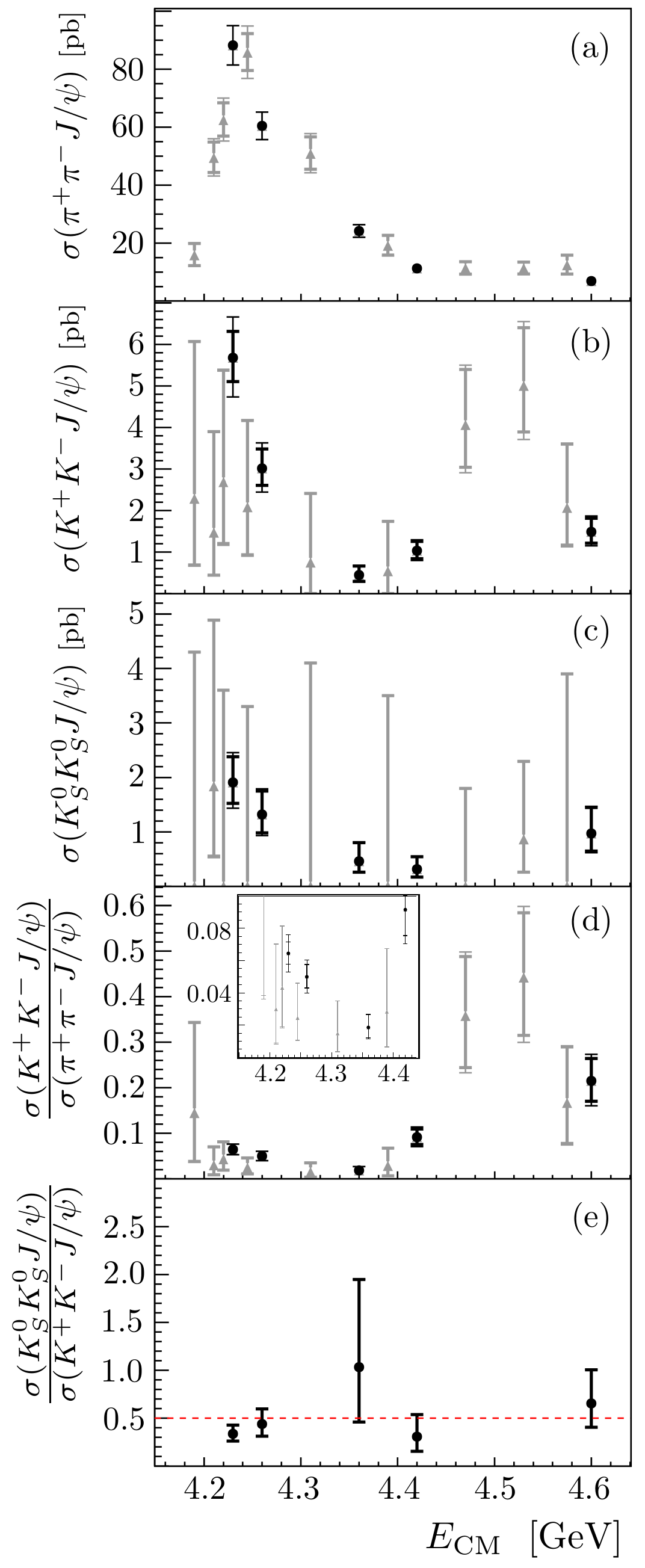}}
\caption{\label{fig:results}
The Born cross sections (a)~$\sige(\pippimjpsi)$, (b)~$\sige(\kpkmjpsi)$, and (c)~$\sige(\ksksjpsi)$, and the ratios (d)~$\sige(\kpkmjpsi)$ / $\sige(\pippimjpsi)$, and (e)~$\sige(\ksksjpsi)$ / $\sige(\kpkmjpsi)$.  
The black circular points are for data sets with high integrated luminosities; the gray triangular points are for smaller data sets.
Thicker error bars are for statistical uncertainties only; thinner error bars are for combined statistical and systematic uncertainties.  In (c), the large error bars with no central point are 90\% C.L. upper limits.  The red dotted line in (e) is the value expected from isospin symmetry.}
\end{figure}

The results for $\sige(\pippimjpsi)$, $\sige(\kpkmjpsi)$, and $\sige(\ksksjpsi)$ are shown in Fig.~\ref{fig:results}(a-c) as functions of $\ecm$ with both statistical and systematic uncertainties.  
To compare the shape of $\sige(\kpkmjpsi)$ with $\sige(\pippimjpsi)$, we calculate the ratio $\sige(\kpkmjpsi)$/$\sige(\pippimjpsi)$~(Fig.~\ref{fig:results}(d)).  If the $Y(4260)$ were the only contribution to the $\pipijpsi$ and $\kkjpsi$ processes, this ratio would be independent of $\ecm$.
This hypothesis is tested by fitting the ratio with a constant for samples with a high integrated luminosity, namely for $\ecm$ of 4.226, 4.258, and 4.358~GeV.  Based on the minimized $\chi^{2}$ of 16.9 with two degrees of freedom and taking into account uncorrelated systematic errors, we find a $3.5\sigma$ standard deviation discrepancy with the assumption of the observed ratio being a constant.  We therefore cannot conclude that the $Y(4260)$ decays through $e^{+}e^{-} \rightarrow \kkjpsi$.

In addition, Fig.~\ref{fig:results}(b) shows a peak near 4.5~GeV in $\sige(\kpkmjpsi)$ that is not present in $\sige(\pippimjpsi)$.  To test the discrepancy between the two channels, we fit $\sige(\kpkmjpsi)/\sige(\pippimjpsi)$ at five $\ecm$ from 4.416 to 4.600~GeV with a constant~(Fig.~\ref{fig:results}(d)).  The resulting $\chi^{2}$ of the fit is $17.6$ for four degrees of freedom, which indicates a $3.0\sigma$ standard deviation discrepancy from the assumption that the ratios are constant.  There is thus evidence for a more complex structure in this region in $\kpkmjpsi$ than in $\pippimjpsi$.

We also calculate the ratios between $\sige(\ksksjpsi)$ and $\sige(\kpkmjpsi)$ for data samples with high luminosity.  According to isospin symmetry, the ratio between these two modes should be $1/2$.  The calculated ratios, along with this prediction, are shown in Fig.~\ref{fig:results}(e). The combined ratio over all energies, based on the total number of signal events, is $0.370^{+0.064}_{-0.058}\pm0.018$, where the first uncertainty is statistical and the second is systematic.

%Final results, including 
%$\sige(\kpkmjpsi)$, $\sige(\ksksjpsi)$, the $\sige(\kpkmjpsi) / \sige(\pippimjpsi)$ ratio, and the $\sige(\ksksjpsi) / \sige(\kpkmjpsi)$ ratio are listed in Table~\ref{table:tabresults}.

Final results are listed in Table~\ref{table:tabresults}.  Upper limits are calculated at a 90\% confidence level and incorporate systematic errors using the Rolke method with an additional uncertainty on the efficiency~\cite{Rolke}.  Systematic uncertainties in the Born cross section measurements are listed in Table~\ref{table:systematics} and are described below. 

%from the integrated luminosity, tracking and PID efficiencies, external branching fractions, $\ks$ reconstruction, kinematic fitting, $J/\psi$ mass resolution, ISR correction, vacuum polarization, $Z_{c}$ substructure, and $K\bar{K}$ substructure.

The integrated luminosity was measured with large-angle Bhabha events and the uncertainty is found to be less than $1\%$~\cite{BES_Lumin}.  
To account for the differences between data and MC simulation in the tracking and PID efficiency, a study was performed using the process $e^{+}e^{-} \to K^{+}K^{-}\pi^{+}\pi^{-}$.  
The systematic uncertainty is found to be $1.0\%$ per charged pion and $2.5\%$ per charged kaon.  The relatively large uncertainty for the charged kaon efficiency is due to the momenta of the charged kaons in this analysis, which are smaller than in typical BESIII analyses.  
For the lepton tracking efficiency, a $1.0\%$ uncertainty per lepton is applied~\cite{etajpsi}.    
We use $J/\psi$ and $\ks$ branching fractions from the PDG~\cite{PDG}, which leads to systematic uncertainties of $0.5\%$. 
The $\ks$ reconstruction efficiency is studied using control samples of $J/\psi \rightarrow \ks K^{\pm}\pi^{\mp}$ and $\phi \ks K^{\pm}\pi^{\mp}$. After factoring out uncertainties due to pion reconstruction and weighting according to the observed $\ks$ momentum distributions, we find a $3.0\%$ systematic uncertainty per $\ks$.

To study the efficiency of the kinematic fit requirements, we used control samples of $e^{+}e^{-} \to \pi^{+}\pi^{-}\pi^{+}\pi^{-}$, $K^{+}K^{-}\pi^{+}\pi^{-}$, and $\ks\ks\pi^{+}\pi^{-}$,
which are similar to $\pippimjpsi$, $\kpkmjpsi$, and $\ksksjpsi$, respectively, but with higher statistics.
Relative efficiencies are defined by comparing yields when requiring $\chi^2$/dof~$<10$ versus $\chi^2$/dof~$<100$. The differences in the efficiencies between MC simulation and data are $2.6\%$ for $\pi^+\pi^-\pi^+\pi^-$, $3.8\%$ for $K^+K^-\pi^+\pi^-$, and $5.9\%$ for $\ks\ks\pi^+\pi^-$, which are taken as the systematic uncertainties.
	
To account for differences in $J/\psi$ mass resolution between data and MC simulation, we smear the width of the $J/\psi$ peak in the signal MC samples by $30\%$.  The changes in the efficiencies of each mode are less than $1.0\%$, which are incorporated as a systematic uncertainty.
	
	The uncertainty associated with the ISR correction factor is studied by replacing the iterative process, described previously, with a $Y(4260)$ Breit-Wigner cross section.
The differences in the Born cross section between these two scenarios are $4.0\%$ for $\pipijpsi$ and $6.0\%$ for $\kkjpsi$, which are taken as the uncertainty for the ISR correction.  Uncertainties on the vacuum polarization corrections are estimated to be $0.5\%$ according to Ref.~\cite{Vac_pol}.

To account for substructure in the $\pipijpsi$ mode,
we compare the efficiency obtained with a phase-space MC sample to that for the process $e^{+}e^{-} \rightarrow \pi^{\pm}Z_{c}(3900)^{\mp} \rightarrow \pippimjpsi$, using the PDG parameters for the $Z_{c}(3900)$.  A $4.0\%$ difference in efficiency is assigned as a conservative systematic uncertainty.  

For the $\kkjpsi$ modes, there is an apparent discrepancy in the $K\bar{K}$ mass spectra between data and MC samples simulated with the phase-space model.  We therefore weight the efficiency according to the observed $\mkpkm$ distribution.  This results in a $10\%$ difference with respect to the nominal efficiency, which is assigned as a systematic uncertainty.

All of these uncertainties are summarized in Table \ref{table:systematics}.  The total systematic uncertainties are $7.5\%$ for $\pippimjpsi$, $14.1\%$ for $\kpkmjpsi$, and $14.5\%$ for $\ksksjpsi$. Taking into account correlations among uncertainties, the systematic uncertainty on the $\sige(\ksksjpsi)/\sige(\kpkmjpsi)$ ratio is 7.2\% and that on the $\sige(\kpkmjpsi)/\sige(\pippimjpsi)$ ratio is 14.6\%.

\begin{table*}[ht]
\caption{\label{table:tabresults}
The center-of-mass energies~($\ecm$), integrated luminosities~($\mathcal{L}$), and final results for $\sige(\kpkmjpsi)$, $\sige(\ksksjpsi)$, $\sige(\ksksjpsi) / \sige(\kpkmjpsi)$, and $\sige(\kpkmjpsi) / \sige(\pippimjpsi)$.  The first uncertainty is statistical, and the second is systematic.  In the cases where there are zero signal events and zero sideband events,
upper limits are 
calculated with 90\% confidence levels and incorporate systematic uncertainties.   The   $\sige(\ksksjpsi) / \sige(\kpkmjpsi)$ ratio is only calculated for data samples with high integrated luminosity.
}
\begin{ruledtabular}
\begin{tabular}{ccccccc}
$\ecm$~[GeV] & $\mathcal{L}$~[pb$^{-1}$] & $\sige(\kpkmjpsi)$ [pb] & $\sige(\ksksjpsi)$ [pb] 
&  \large{$\frac{\sige(\ksksjpsi)}{\sige(\kpkmjpsi)}$} 
& \large{$\frac{\sige(\kpkmjpsi)}{\sige(\pippimjpsi)}$}\\ 
\hline
4.189 &
43 %43.09LUM
 &
 $2.2^{+3.8}_{-1.6}\pm0.3$ & $ <4.3 $ 
 &$ - $ & $0.14^{+0.20}_{-0.10}\pm0.02$ \\ 
4.208 &
55 %54.55LUM 
&
 $1.4^{+2.4}_{-1.0}\pm0.2$ & $ 1.7^{+3.0}_{-1.3}\pm0.3 $ 
& $ - $ & $0.030^{+0.042}_{-0.021}\pm0.004$ \\ 
4.217 &
54 %54.13LUM 
&
  $2.5^{+2.7}_{-1.5}\pm0.4$ & $ <3.6  $ 
 &$ - $ & $0.043^{+0.037}_{-0.022}\pm0.006$ \\ 
4.226 &
1092 %44.40+1047.34LUM 
&
 $5.27^{+0.63}_{-0.57}\pm0.75$ & $ 1.6^{+0.5}_{-0.4}\pm0.3 $ 
 &$0.307^{+0.090}_{-0.072}\pm0.024$ & $0.0644^{+0.0067}_{-0.0062}\pm0.0094$ \\ 
4.242 &
56 %55.59LUM 
&
  $2.0^{+2.1}_{-1.1}\pm0.3$ & $ <3.3 $ 
& $ - $ & $0.024^{+0.023}_{-0.017}\pm0.004$ \\ 
4.258 &
826 %523.74+301.93LUM 
&
 $3.08^{+0.47}_{-0.41}\pm0.40$ & $ 1.2^{+0.4}_{-0.3}\pm0.2 $ 
& $0.40^{+0.15}_{-0.12}\pm0.04$ & $0.0499^{+0.0082}_{-0.0074}\pm0.0073$ \\ 
4.308 &
45 %44.90LUM 
&
 $0.7^{+1.7}_{-0.7}\pm0.1$ & $ <4.1  $ 
 &$ - $ & $0.015^{+0.026}_{-0.014}\pm0.002$ \\ 
4.358 &
540 %539.84LUM 
&
 $0.43^{+0.22}_{-0.15}\pm0.06$ & $ 0.44^{+0.34}_{-0.20}\pm0.07 $ 
 & $1.03^{+1.01}_{-0.56}\pm0.08$ & $0.0185^{+0.0083}_{-0.0065}\pm0.0027$ \\ 
4.387 &
55 %55.18LUM 
&
 $0.4^{+1.2}_{-0.4}\pm0.1$ & $ <3.5  $ 
 & $ - $& $0.028^{+0.050}_{-0.024}\pm0.004$ \\ 
4.416 &
1074 %44.67+1028.89LUM 
&
 $0.97^{+0.22}_{-0.19}\pm0.14$ & $ 0.34^{+0.23}_{-0.15}\pm0.05 $ 
 &  $0.35^{+0.24}_{-0.15}\pm0.02$ &  $0.091^{+0.019}_{-0.017}\pm0.013$\\ 
4.467 &
110 %109.94LUM 
&
 $3.8^{+1.3}_{-1.0}\pm0.5$ & $ <1.8  $ 
 & $ - $ & $0.36^{+0.15}_{-0.11}\pm0.05$ \\ 
4.527 &
110 %109.98LUM 
&
 $4.3^{+1.4}_{-1.1}\pm0.7$ & $ 0.82^{+1.43}_{-0.60}\pm0.13 $ 
 & $ - $ & $0.44^{+0.15}_{-0.11}\pm0.06$ \\ 
4.575 &
48 %47.67LUM 
&
  $2.0^{+1.5}_{-0.9}\pm0.3$ & $<3.9  $ 
 & $ - $& $0.17^{+0.12}_{-0.07}\pm0.02$ \\ 
4.600 &
567 %566.93LUM 
&
 $1.42^{+0.33}_{-0.27}\pm0.20$ & $ 0.92^{+0.50}_{-0.35}\pm0.14 $ 
& $0.65^{+0.36}_{-0.25}\pm0.05$ & $0.215^{+0.052}_{-0.043}\pm0.031$ \\ 	
\end{tabular}
\end{ruledtabular}
\end{table*}

\begin{table}[ht]
\caption{\label{table:systematics}
Summary of systematic uncertainties. }
\begin{ruledtabular}
\begin{tabular}{cccc}
 & $\pippimjpsi$ & $\kpkmjpsi$ & $\ksksjpsi$\\ 
\hline
Luminosity& $1.0\%$ & $1.0\%$ & $1.0\%$ \\ 
Tracking and PID& $4.0\%$ & $7.0\%$ & $6.0\%$ \\ 
Branching Ratios& $0.5\%$ & $0.5\%$ & $0.5\%$ \\ 
$\ks$ Reconstruction & - & - & $6.0\%$ \\ 
$J/\psi$ Resolution& $1.0\%$ & $1.0\%$ & $1.0\%$ \\ 
Kinematic Fit& $2.6\%$ & $3.8\%$ & $5.9\%$ \\ 
Vacuum Polarization & $0.5\%$ & $0.5\%$  & $0.5\%$ \\ 
ISR Correction& $4.0\%$ & $6.0\%$ & $6.0\%$ \\ 
$Z_{c}$ Substructure & $4.0\%$ & - & - \\ 
$KK$ Substructure& - & $10.0\%$ & $10.0\%$ \\ 
\hline
Total& $7.5\%$ & $14.1\%$ & $14.5\%$ \\ 
\end{tabular}
\end{ruledtabular}
\end{table}

In summary, we measure the Born cross sections as functions of $\ecm$ for the processes $e^{+}e^{-} \rightarrow \kpkmjpsi$, $\ksksjpsi$, and $\pippimjpsi$.  
We also measure the ratios of Born cross sections for $\ksksjpsi$ to $\kpkmjpsi$ and $\kpkmjpsi$ to $\pippimjpsi$. 
The results suggest the 
$\kpkmjpsi$ and $\pippimjpsi$ cross sections have different energy dependencies in the region around the $Y(4260)$. 
In addition, there is evidence for an enhancement in the cross section of $e^{+}e^{-} \rightarrow \kkjpsi$ in the higher $\ecm$ region.  More data and additional analyses are needed to investigate the nature of this structure.
We find the ratio of cross sections for the reactions with neutral and charged kaons to be consistent with expectations from isospin conservation.

The BESIII collaboration thanks the staff of BEPCII and the IHEP computing center for their strong support. This work is supported in part by National Key Basic Research Program of China under Contract No. 2015CB856700; National Natural Science Foundation of China (NSFC) under Contracts Nos. 11235011, 11322544, 11335008, 11425524, 11635010; the Chinese Academy of Sciences (CAS) Large-Scale Scientific Facility Program; the CAS Center for Excellence in Particle Physics (CCEPP); the Collaborative Innovation Center for Particles and Interactions (CICPI); Joint Large-Scale Scientific Facility Funds of the NSFC and CAS under Contracts Nos. U1232201, U1332201, U1532257, U1532258; CAS under Contracts Nos. KJCX2-YW-N29, KJCX2-YW-N45; CAS Key Research Program of Frontier Sciences under Contract No. QYZDJ-SSW-SLH003; 100 Talents Program of CAS; National 1000 Talents Program of China; INPAC and Shanghai Key Laboratory for Particle Physics and Cosmology; German Research Foundation DFG under Contracts Nos. Collaborative Research Center CRC 1044, FOR 2359; Istituto Nazionale di Fisica Nucleare, Italy; Koninklijke Nederlandse Akademie van Wetenschappen (KNAW) under Contract No. 530-4CDP03; Ministry of Development of Turkey under Contract No. DPT2006K-120470; National Science and Technology fund; The Swedish Research Council; U. S. Department of Energy under Contracts Nos. DE-FG02-05ER41374, DE-SC-0010118, DE-SC-0010504, DE-SC-0012069; University of Groningen (RuG) and the Helmholtzzentrum fuer Schwerionenforschung GmbH (GSI), Darmstadt; WCU Program of National Research Foundation of Korea under Contract No. R32-2008-000-10155-0.

%\end{linenumbers}

\end{document}